\documentclass[footinbib, aps,prl,twocolumn]{revtex4}
\usepackage{graphicx}
\usepackage{dcolumn}
\usepackage{amsmath}
\usepackage{amssymb}
\usepackage{float}
\usepackage{epstopdf}
\usepackage{color}
\usepackage{soul}

\def\kv{{\bf k}}

\def\beq{\begin{equation}}
\def\eeq{\end{equation}}
\def\beqa{\begin{eqnarray}}
\def\eeqa{\end{eqnarray}}

\begin{document}

\title{Photon Inhibited Topological Transport in Quantum Well Heterostructures}
\author{Aaron Farrell and T. Pereg-Barnea}
\affiliation{Department of Physics and the Centre for Physics of Materials, McGill University, Montreal, Quebec,
Canada H3A 2T8}
\date{\today}
\begin{abstract}
Here we provide a picture of transport in quantum well heterostructures with a periodic driving field in terms of a probabilistic occupation of the topologically protected edge states in the system. This is done by generalizing methods from the field of photon assisted tunneling.  We show that the time dependent field  {\it dresses} the underlying Hamiltonian of the heterostructure and splits the system into side-bands. Each of these sidebands is occupied with a certain probability which depends on the drive frequency and strength. This leads to a reduction in the topological transport signatures of the system because of the probability to absorb/emit a photon. Therefore when the voltage is tuned to the bulk gap the conductance is smaller then the expected $2e^2/h$.
We refer to this as photon inhibited topological transport. Nevertheless, the edge modes reveal their topological origin in the robustness of the edge conductance to disorder and changes in model parameters.
In this work the analogy with photon assisted tunneling allows us to interpret the calculated conductivity and explain the sum rule observed by previous authors\cite{Kundu}
\end{abstract}
\maketitle

\emph{Introduction.}---Topological states of matter are currently at the forefront of research in condensed matter physics. From the quantum hall effect to topological superconductors, these states are of interest for a variety of reasons. In topological insulators the in-gap edge states are of primary interest. These states are topologically protected, meaning they are insensitive to deformations of the Hamiltonian's parameters that leave the topological gap intact and the effects of disorder. The existence of such states provides a physical signature of the topology in the charge and spin conductance.

Recently, there has been a growing amount of attention paid to the generation and/or manipulation of topological states of matter through the application of a time-periodic perturbation\cite{Lindner, Gu,Oka,Usaj, Calvo,Torres, Leon, Rudner, Kitagawa2,Kundu2,Katan, Jiang, Kundu, Liu, Wu, Wang2, Delplace, Li4, paraj, karthik}. Experimental progress in this direction has been made in both photonic crystals\cite{rechtsman} and in a solid-state context in Bi$_2$Se$_3$\cite{Wang}. In the Letter we study how a time-periodic perturbation can be used to manipulate the transport properties of a quantum spin Hall insulator. For example: such a system is expected to have a two terminal conductivity of $2e^2/h$ in equilibrium. With the application of a time-periodic field, we find that this value may be reduced significantly. Despite this reduction and a deviation from quantized units of $e^2/h$, we find that this conductivity is still topological in the sense that it is robust to disorder, system size changes, and gap-conserving deformations of the Hamiltonian. Furthermore, we describe a method to predict the degree of these deviations quantitatively, and their dependence on the drive strength and frequency.

To understand how this reduction in the conductivity can be tuned and why it appears to be topologically robust,  we have developed an understanding by generalizing the viewpoint of photon assisted tunneling\cite{TienGordon, Platero20041}. We find that the periodic perturbation has a two-pronged effect. First, it ``dresses" the original static Hamiltonian and second, it causes the edge conductance channels to only be occupied probabilistically upon the injection of a lead electron. This is because electrons tunneling into the system can absorb/emit a photon. In this sense, the presence of the photons inhibit the topological transport properties of the system. This description not only accounts for the reduction of the conductivity, but also explains why its values are topological in nature. This interpretation will be important for transport experiments in Floquet topological insulators. It provides an explanation of why the conductivity isn't quantized as well as shows that the conductivity can potentially be tuned predictably in the lab.

\emph{Methods.}--- As a model system we take the quantum well heterostructures who play host to the quantum spin Hall effect. We apply a time dependent field and allow for on-site disorder. Our Hamiltonian is as follows
\beqa
H_S=H_{QW}+H_{disorder}+H_{ext}(t)
\eeqa
where $H_{QW}=\sum_{\kv} \psi_\kv^\dagger \left(   \begin{matrix}   \hat{ H}(\kv) & 0 \\   0 &  \hat{ H}^*(-\kv)\\ \end{matrix}\right) \psi_\kv$ where $\psi_\kv^\dagger$ is a four component creation operator for electrons at momenta $\kv$ in state $m_J=(1/2,3/2,-1/2,-3/2)$ of the clean heterostructure and $\hat{ H}(\kv)= \epsilon_\kv \sigma_0 + {\bf d}(\kv)\cdot {\bf \sigma}$, with $\sigma$ being a vector of Pauli matrices. In the typical language of these structures\cite{Bernevig, konig}, we take ${\bf d}(\kv)=(A\sin{k_x},A\sin{k_y}, M-4B+2B(\cos{k_x}+\cos{k_y}))$ and $\epsilon_\kv =C-2D(2-\cos{k_x}-\cos{k_y})$. In order to focus on transport without additional complications we follow Lindner and coworkers\cite{Lindner} and set $C=D=0$, $A=B=0.2|M|$. All energies are in units of $M$. As we are interested in a ``topological" system we take $M=1$ so that $\text{sgn}(M/B)=1$\cite{Lindner,Bernevig} .

Next, $H_{ext}(t) =2 ({\bf V}\cdot \sigma) \cos{\Omega t}$ is an electromagnetic field polarized in the direction ${\bf V}$[\onlinecite{Ding2014,ZhangTHZ,Lindner, Katan}]. For concreteness, we will take $ {\bf V}= V_{ext} \hat{z}$; although this is not necessary for what follows.  {Note $H_{ext}(t)$ obeys the periodic generalization of time-reversal invariance\cite{Lindner}} $\mathcal{T} H_{ext}(t) \mathcal{T} ^{-1}= H_{ext}(-t+\tau)$ for some $\tau$. Finally, $H_{disorder}=-\sum_{i,\alpha} w_i \psi^\dagger_{i,\alpha} \psi_{i,\alpha}$ ( with $ \psi_i^\dagger$ the Fourier transform of $ \psi_\kv^\dagger$). This corresponds to charge impurities (disorder) changing the chemical potential on each site by $w_i$. We draw the $\{w_i\}$ randomly from an evenly distributed sample between $-W/2$ and $W/2$. We call $W$ the disorder strength.

 \begin{figure}[tb]
  \setlength{\unitlength}{1mm}
     \includegraphics[scale=.4]{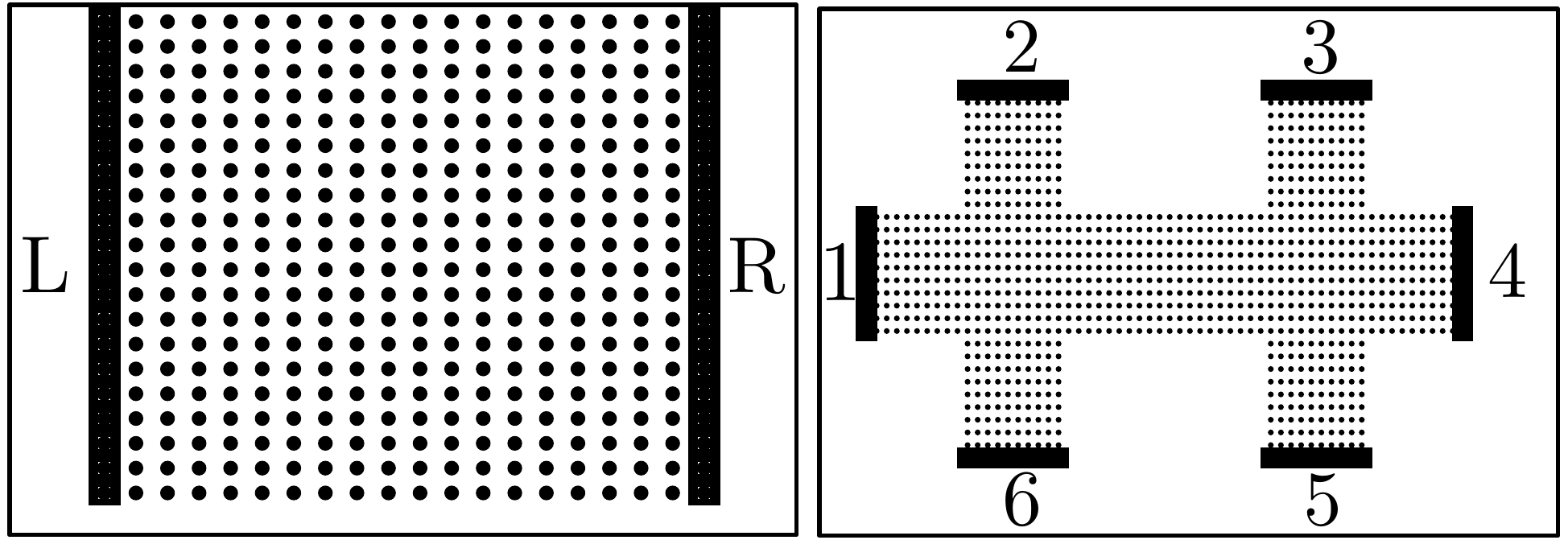}
\caption{{\small
The two device geometries considered in this work. Left is a two-terminal device labeled with leads left (`L'), and right (`R'). On the right is a six terminal device labeled with leads 1through 6 . The sites coupled to leads have a solid rectangle around them.
     }
     }\label{fig:devices}
\end{figure}

Our numerical study employs the Floquet-Landauer formalism\cite{Kohler, Kitagawa, Kundu, Torres}.  Similar to Ref.~\cite{Torres}, we consider two different device geometries ( see Fig. \ref{fig:devices}). First, we consider a two-terminal device with the left and right end of the system attached to leads whose Fermi level lies at the "lead energy" $E$ with a slight offset bias between the two leads \cite{Supp}\nocite{Martinez, Martinez2}.  {In this set-up the quantity of interest is $\sigma(E)$, the differential conductivity given that the chemical potential of the leads is at energy $E$}. For a spin Hall insulator ({\it e.g.} our model above) in equilibrium when the lead energy $E$ in a two-terminal device is tuned to lie in the gap ({\it i.e.} on the edge states), a value $\sigma(E)=2e^2/h$ is expected\cite{Tworz, Kundu2}. This is the first signature in which we are interested. For convenience, we define $\sigma_{TT}=\sigma(V\simeq0)h/e^2$. Secondly, we consider a six-terminal device. This device allows us to probe whether the current is carried by bulk or edge modes\cite{Chen, Roth, Torres}. In equilibrium, it is found that the only non-zero values of the transmission elements between leads $\lambda$ and $\lambda'$, $T_{\lambda, \lambda'}(\epsilon)$ (with $\epsilon$ in the gap), come from tunneling between adjacent leads in the device. Thus $T_{\lambda, \lambda'}(\epsilon_F)=0$, unless $\lambda = \lambda'\pm1$ (where $6+1\to1$). Moreover, it is argued that $T_{\lambda, \lambda\pm1}(\epsilon_F)=1$ as, because of the helical edge states, a quasiparticle originating at lead $\lambda$ {\em must} tunnel to one of the neighbouring leads. Later in this Letter we look for similar properties in the non-equilibrium system.

Before proceeding we comment on recent criticisms of Floquet states in periodically driven systems\cite{DAlessio, DAlessio2, PhysRevE.90.012110, Ponte}. Floquet states are often thought of as the steady-states of a time-periodic system\cite{Sambe}. Refs. \cite{DAlessio, PhysRevE.90.012110, Ponte} argue that the long time evolution of an isolated,  periodically driven system leads to an effectively infinite temperature state for some driving periods.  Our formalism for calculating transport properties attaches leads to the system (i.e. it is not isolated anymore) and only makes assumptions about the state of the leads in the distant past, namely that the leads are in a thermal equilibrium and no assumptions on the state of the system\cite{Supp}. This assumption provides the state of the system at the present time and does not rely on ``evolving" any particular Floquet state.

\begin{figure}[tb]
  \setlength{\unitlength}{1mm}
     \includegraphics[scale=.45]{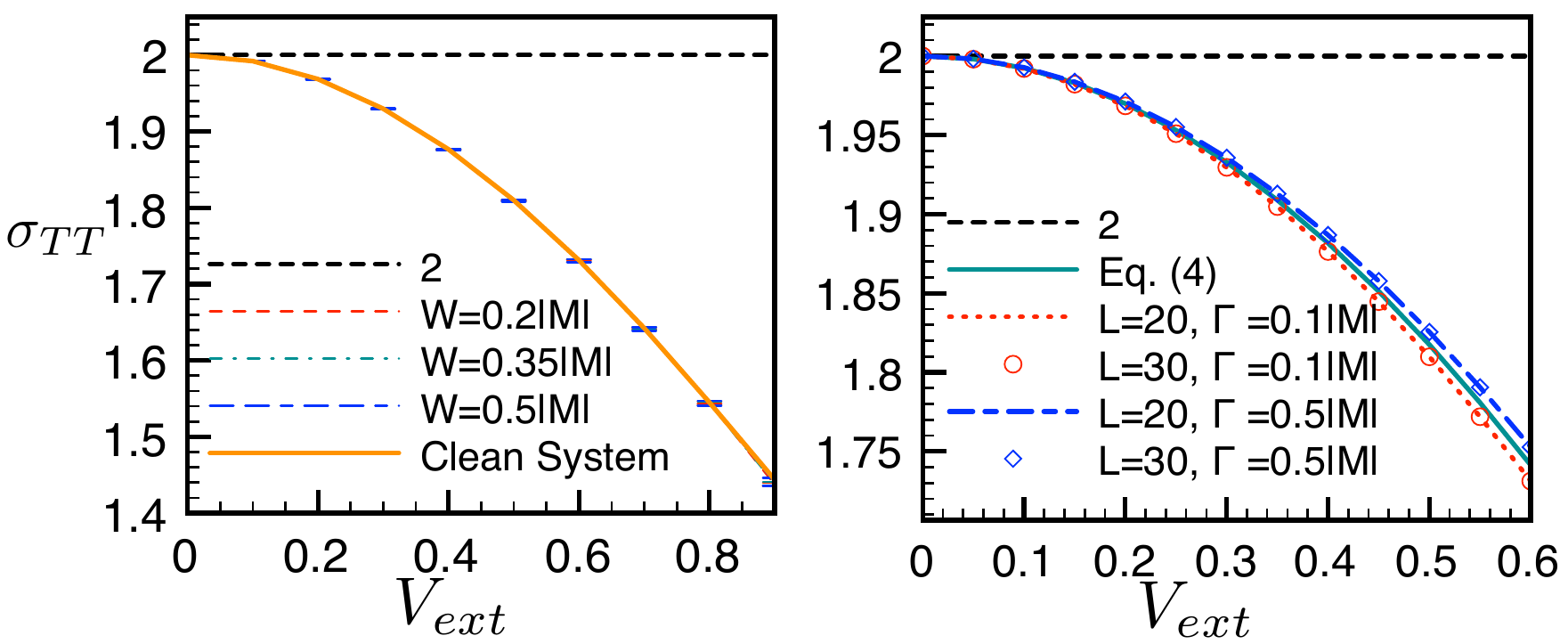}

\caption{\small{\small
Plots of the differential two-terminal conductivity as a function of $V_{ext}$. Left: results for various disorder strengths, right: various values of the system size ($L$) and the lead coupling parameter ($\Gamma$).   }
     }\label{fig:TopSystem}
\end{figure}

\emph{Transport Results.}--- We begin with a clean system ($W=0$) in a two terminal geometry.  We fix $\Omega = 2.3|M|$, and tune $V_{ext}$. We plot $\sigma_{TT}$  for $V_{ext}=0|M|...|M|$ in Fig. \ref{fig:TopSystem}. As $V_{ext}$ is increased from zero, the quantization of $\sigma_{TT}$ is lost. For moderately strong $V_{ext}$, we see that it reaches $\sigma_{TT}\sim 1.5 $. This shows that for a quantum spin Hall insulator, the (bare) conductivity is {\it not}, in general, quantized to the traditional equilibrium value under the application of a periodic perturbation.

Looking again at Fig.~\ref{fig:TopSystem}, we see that these values are robust to the strength of the disorder potential. The deviation from the clean limit is insignificant, even up to disorder strengths of $M/2$. Additionally, these values are insensitive to the coupling strength of the system to the leads\cite{Supp}, ${\Gamma}$, and the system size. This robustness leaves the impression that despite the deviation of the conductivity from $\sigma_{TT}=2$, the values it takes appear to be topologically protected. Our six-terminal calculations provide additional evidence of topological, edge conductance. With the lead energy set in the gap of the system, we find that  $T_{\lambda,\lambda'}=0$, except the off-diagonal elements $T_{\lambda, \lambda+1}$ and $T_{\lambda+1,\lambda}$. In contrast to equilibrium, we find that $T_{\lambda, \lambda+1}=T_{\lambda+1,\lambda}<1$. In spite of this, we observe that the conduction takes place {\em only} between adjacent leads suggesting that the current is only flowing on the edges.

{To explain the above behavior, we borrow insight from the field of photon assisted tunneling (PAT). PAT, as first proposed by Tien and Gordon\cite{TienGordon}, was originally used to describe a superconducting-insulator-superconductor tunnel junction. When a periodic AC voltage $V_{ac}$ is applied to one of the leads, the energy eigenstates of these leads split into sidebands at energy $E+n\hbar\Omega$ for integer $n$ and driving frequency $\Omega=2\pi/T$. The probability that each one of these side bands is occupied is given by $J_n^2(\alpha)$, where $\alpha=eV_{ac}/\hbar\Omega$, and $J_n$ is the $n^{th}$ Bessel function of the first kind. The consequence of this sideband splitting is that when a lead energy $E$, is applied across the tunnel junction, the electrons can tunnel into the system not just at energy $E$, but at $E+n\hbar\Omega$ with a probability of $J_n^2(\alpha)$. One interprets this as the electrons absorbing ($n>0$) or emitting ($n<0$) $|n|$ photons. As a result the conductivity in the driven system is given by $\sigma_{PAT}(E)= \sum_{n}  J_n^2(\alpha)\sigma_ {0}(E+n\hbar\Omega)$\cite{TienGordon, Platero20041}. Here $\sigma_ {0}(E)$ is the conductivity of the junction in the absence of the AC voltage. }

Here we do not have a simple periodic modulation of the sample system, rather the modulation itself has some internal structure given by ${\bf V}\cdot \sigma$. The result of this is that the system is not simply split into side-bands. The fact that $H_{ext}(t)$ does not commute with the static Hamiltonian, leads to interesting effects. In the case of off-resonant light (light where $\hbar\Omega$ does not connect parts of the clean, static spectrum), we can make some simplifying assumptions to obtain an effective description in line with PAT. We describe this simpler case here and leave the discussion of on-resonant light, where more care must be taken, for later\cite{Farrelltriv}.

In the field of Floquet topological insulators \cite{Jiang, Kundu, Liu, Wu,Kitagawa} with off-resonant light, it is known that one can think of the periodic perturbation as``dressing" the static system by modifying its underlying physical parameters to produce a new, effective {\em static} Hamiltonian. However, this approach is incomplete from a transport point of view. One must take into account the splitting of the states of this effective Hamiltonian into side-bands. Thus off-resonant light has a two-sided effect: First, it dresses the static Hamiltonian to produce a new effective static Hamiltonian. Second, the eigenstates of this effective Hamiltonian are split into side-bands in a process analogous to PAT.  {This picture is not specific to the illustrative system we have chosen here, it is more general. It may be used, for example, to describe transport calculations in analogous systems like illuminated graphene}. 

To motivate this consider writing $|\psi(t)\rangle=U_V(t)|\hat{\psi}(t)\rangle$ where $i\hbar \frac{d }{dt}U_V(t)=H_{ext}(t)U_V(t)$. This transforms our problem into a new problem with the Hamiltonian $\hat{{H}}(t)=U^\dagger_V(t)H U_V(t)$, where $H$ is the Hamiltonian in the absence of the time-dependent field. If $[H_{ext}(t), H]=0$ then $ \hat{{H}}(t)= H$ leading to an analogue of traditional PAT. Here $[H_{ext}(t), H]\ne0$ in general and the transformation $U_V(t)$ leads to a new, time-dependent Hamiltonian. However, provided the mixing between bands is weak (off-resonant), it is possible to approximate $\hat{{H}}(t)$ by its time-averaged value. In the language of Floquet theory this amounts to the leading order term\cite{Kitagawa} in $\hat{{H}}(t)$ of the Floquet Hamiltonian $H_F=\frac{i}{T}\text{T}\left(e^{-i\int_0^T dt\hat{{H}}(t)}\right)$, $\text{T}(\cdot)$ denoting time ordering.

One can study the transport properties of this new effective Hamiltonian. This, however, will miss the unitary transformation that we have performed to get this Hamiltonian. Accounting for this transformation in a full transport calculation, in the approximation described above, we arrive at the following expression for the two-terminal conductivity of this system\cite{Supp}
\beq\label{PATsigma}
\sigma(E)= \sum_{m} J_{m}^2\left(\frac{2V_{ext}}{\hbar\Omega}\right) \sigma_F(E+m\hbar\Omega)
\eeq
where $\sigma_F(E)$ is the {\em static} differential conductivity of the dressed system described by $H_F$. For our current model, we have a finite band width and have not taken into account higher (or lower) energy bands. We assume the bands near the Fermi level are separated in energy from the other bands by a sufficient amount so that they can be neglected. Experimental validation of this comes from Ref.~\cite{Wang} where the experimental results can be understood by using only the bands near the Fermi level. As a result we have $\sigma_F(E+m\hbar\Omega)=\sigma_F(V)\delta_{m,0}$ for $E\simeq0$; no states exist at $m\hbar\Omega$. Therefore, we have
\beq\label{Besselcompare}
\sigma(E)=J_{0}^2\left(\frac{2V_{ext}}{\hbar\Omega}\right) \sigma_F(E) \ \ \ (E\simeq0)
\eeq
Thus with an off resonant driving frequency, we describe the underlying system with an effective static Hamiltonian which may give rise to the signature transport properties. In the present case, we are interested in a Hamiltonian showcasing  the quantum spin Hall effect. This state should have a two-terminal conductance of $2e^2/h$, and six-terminal transmission elements as described above. In the presence of a driving field, the in-gap edge states are only occupied with a certain probability due to the prospect of absorption/emission of photons. Thus, the transport property we are interested in only shows up with a certain probability. In the present case we expect $\sigma_F(V) =2e^2/h$, and so the actual conductivity we measure will be $\sigma(E)=2J_{0}^2\left(\frac{2V_{ext}}{\hbar\Omega}\right)$. Plotting this against our numerical data produces excellent agreement (see Fig. \ref{fig:TopSystem}).  {One may look at this expression as a correction to the quantized value of $2e^2/h$. One can show for in-gap energies $E$ that $\sigma(E)\simeq 2\left(1- \left(\frac{V_{ext}}{\hbar\Omega}\right)^2\right)$, {\it i.e.} this correction is second order in $V_{ext}/\hbar\Omega$.}

This explains our observation in the opening of this section. Despite the fact that we do not obtain the values $\sigma=2e^2/h$, or $T_{\lambda, \lambda\pm1}=1$ , the values that we do see are robust in the same way as the equilibrium values. The underlying system is topological in nature, with helical edge states that give rise to $2e^2/h$ conductance and $T_{\lambda, \lambda\pm1}=1$. However there is only a certain probability that the electrons tunneling from the leads are at the correct energy to take advantage of these channels. Thus, the presence of these photons in the system inhibits the ability of these edge channels to transport charge.

\begin{figure}[]
  \setlength{\unitlength}{1mm}

   \includegraphics[scale=.3125]{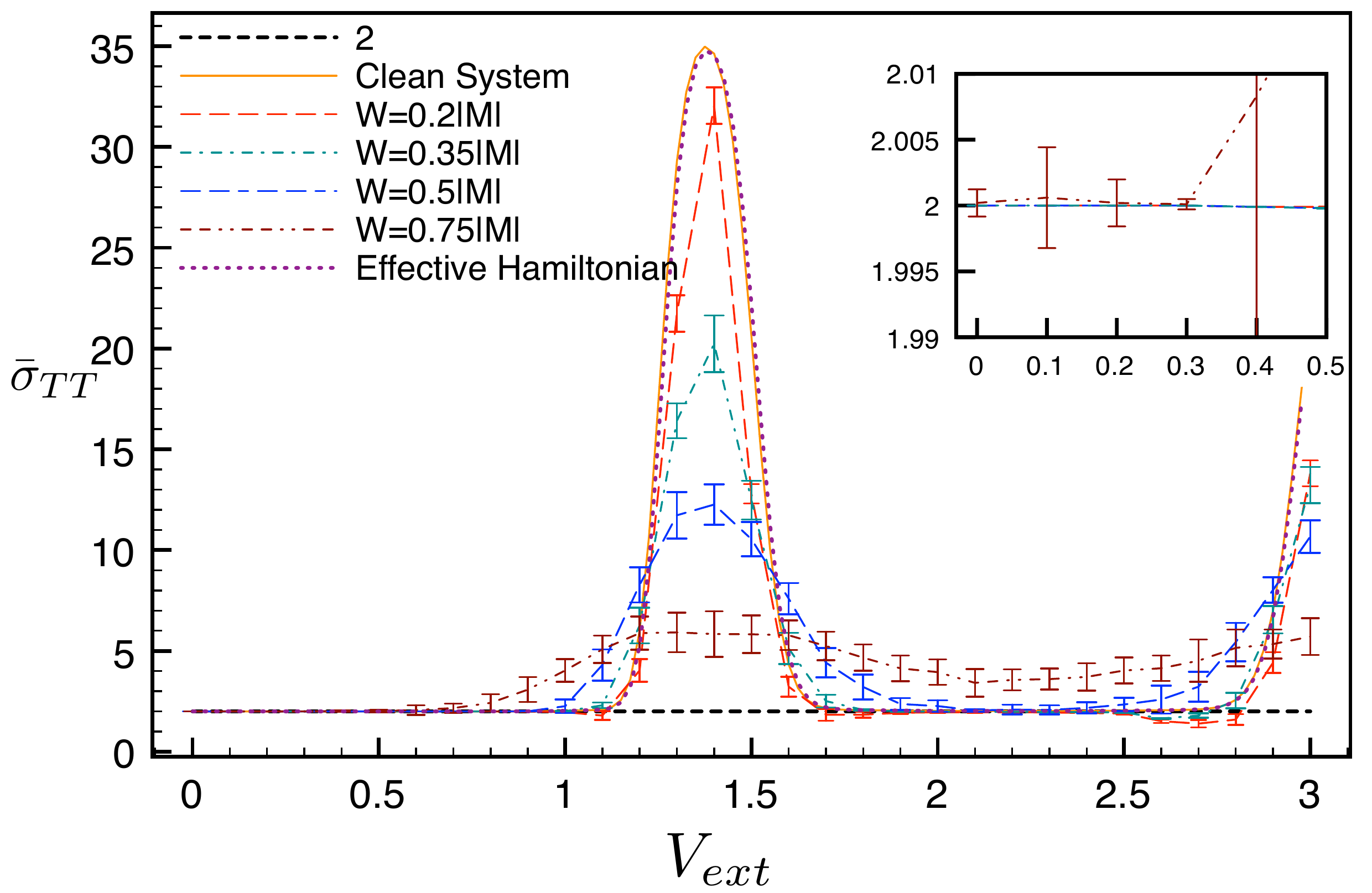}
\caption{{\small
Disorder averaged summed conductivity, Eq. (\ref{sumrule}).  The inset shows a zoomed in picture of the first area of conductivity quantization. Note some error bars in the insets are too small to see.
     }
     }\label{fig:sumrule}
\end{figure}

Our discussion so far has not relied on the fact that the original Hamiltonian is topological in nature, rather it is enough that the effective Hamiltonian be topological. In other systems, it is possible to drive topological states in otherwise trivial systems with off-resonant light. The most prevalent example of this is graphene, where the light produces an effective Hamiltonian with a topological mass. Thus the suppression described above may also apply to these other systems. \cite{Gu,Oka,Usaj, Calvo,Torres, Kundu2, Kitagawa}. In the present system of interest the driving of a trivial equilibrium system ({\it i.e.} $M=-1$ in our current language) can be driven into a topological phase.  This, however, relies on the light being {\em on resonance}\cite{Lindner}. A description of this scenario inline with the discussion above is possible, but subtle and we leave it to a future communication\cite{Farrelltriv}.

\emph{Connection to Floquet Sum Rule.}---We now connect our work to a sum rule proposed recently by Kundu and Seradjeh in the context of a system with Floquet Majorana modes\cite{Kundu}. Similar to the current work, these authors find that in the presence of a periodic perturbation, a system with Majorana modes will not showcase the expected zero-bias quantized conductance of $2e^2/h$. Instead, the quantized conductivity is found in the sum
\beq\label{sumrule}
\bar{\sigma}(E)=\sum_n \sigma(E+n\hbar\Omega).
\eeq
 {Physically, the above corresponds to performing measurements of $\sigma(E)$ not just at an in gap energy $E$, but for lead energies placed any number of $\hbar\Omega$'s above or below this. The results of these measurements are than summed up}. 
Let us apply this sum rule to our system. Using Eq.~(\ref{PATsigma}) we have
\begin{equation}
\bar{\sigma}(E)=\sum_{n,m} J_{m}^2\left(\frac{2V_{ext}}{\hbar\Omega}\right) \sigma_F(E+(m+n)\hbar\Omega).
\end{equation}
Shifting $n\to n-m$, using the off resonance light conductivity ,$\sigma_F(E+m\hbar\Omega)=\sigma_F(E)\delta_{m,0}$ and the Bessel functions property $\sum_n J_n^2(x)=1$ leads to $\bar{\sigma}(E)= \sigma_F(E)$. Therefore, if $\sigma_F(E)$ is quantized to $2e^2/h$, then $\bar{\sigma}(E)$ should be as well.

The above result is intuitive from a PAT point of view. At a two-terminal lead energy $E\simeq n\hbar\Omega$ the electrons must emit $n$ photons to enter the quantized conductance channel and thus enter it with probability $P_{-n}$, the probability to emit $n$ photons. This gives a conductance of $\sigma_F(0)P_{-n}$.  Summing over all the lead energies is then effectively summing over all of the probabilities as $\bar{\sigma}(0)=\sigma_F(0)  \sum_n P_{n} = \sigma_F(0)$, {\it i.e.} the sum rule recovers the underlying conductance.

{The above derivation can be generalized to on-resonant driving under certain conditions\cite{Farrelltriv}. In particular, one expects the sum rule to hold when edge states are visible in the so-called "quasi-energy" spectrum. Nonetheless, the derivation presented here contains all of the intuition required to understand the sum rule.}

In Fig.~\ref{fig:sumrule} we show $\bar{\sigma}(E)$ at $E=0$ for various different disorder strengths as well as $\sigma_F(E)$. Firstly, our data for the clean system is in excellent agreement with $\sigma_F(E)$. Secondly, the system shows noticeable deviations from $\bar{\sigma}(E\simeq0)=2e^2/h$ in two regimes of $V_{ext}$ and occur in both the clean and disordered systems. Here the bulk gap in the effective Hamiltonian closes, and the topological edge states becoming washed out by bulk conduction states. This is most obvious when looking at the disorder averaged data where the regions with $\bar{\sigma}(E)=2e^2/h$ are insensitive to disorder, while the peaks are sensitive to disorder, as bulk conduction states should be. This result is interesting from a PAT perspective. Not only has the periodic field split the system into side bands, but it has modified the underlying system in a non-trivial way. In a traditional PAT context only the sideband splitting would take place.

\emph{Conclusions.}---We have developed an analogue of PAT to describe the transport signatures of topologically protected edge states in the quantum well heterostructures. Our picture entails electrons only accessing the topological edge states of the system probabilistically. The probability of the electrons to absorb/emit a photon reduces the traditional values associated with transport measurements in these systems. These reduced values are, however, still insensitive to disorder and other deformations. We refer to this phenomenon as ``photon inhibited topological transport".

By using this picture we related our system to a Floquet sum rule proposed before\cite{Kundu}. Our picture of PAT is able to offer a physical description of why one would expect such a rule to hold. Namely, the sum rule is adding up all of the probabilities of accessing the edge state which, by itself, should have the traditional transport signatures. This sum then reveals the underlying transport properties.

\begin{acknowledgements}

The authors are thankful for useful discussions with Jean-Ren\'e Soquet, Aashish Clerk and Gil Refael. Financial support for this work was provided by the NSERC and FQRNT (TPB) and the Vanier Canada Graduate Scholarship (AF). Numerical calculations for this work were performed using McGill HPC supercomputing resources.

\end{acknowledgements}

\bibliographystyle{apsrev}
\bibliography{Floquet}
\footnotetext{Footnote `1'\label{fnt:1}.}

\end{document}